\documentclass{article}
\usepackage{amssymb}
\usepackage{amsmath}
\usepackage{harvard}

\input{tcilatex}
\begin{document}

\title{Ghost- Free Higher Derivative Quantum Gravity, the Hierarchy and the
Cosmological Constant Problems }
\author{V. I. Tkach \\
Department of Physics and Astronomy,\\
Northwestern University, Evanston, IL 60208-3112, USA\\
\textit{v-tkach@northwestern.edu}}
\maketitle

\begin{abstract}
Proposing a new solution to problems of the hierarchy and smallness

of the cosmological constant using the Tev scale of the Standard Model

in a new framework of the higher-order gravity.
\end{abstract}

\begin{center}


\bigskip

\ \ \ \ \ \ 
\end{center}

The theory of higher derivative gravitation, whose action contains terms
quadratic in the curvature in addition to the Einstein term, is a
renormalizable field theory $^{1-7}$, but it is not free defect. This theory
gives rise to unphysical poles in spin-two sector of the tree- level
propagator which break the unitarity.

On the other hand, induced gravity program with fourth --order gravitational
theories $^{8-12}$ that does not contain dimensional coupling constants and
the unphysical ghosts, but in such theories Newton's constant is not
calculable and is a free parameter $^{13}$.

In this note we present an example of the quantum gravity with the higher
curvature which is ghost-free.

We show that the gravitational strength and other observed fundamental
interactions are the consequence of one fundamental dimension scale about $%
10^{3}\ Gev$ , which is vacuum expectation values the Higgs fields $v$=$%
M_{\varepsilon w}\approx 10^{3}Gev$ of the Standard Model. This is a new
solution to the well-known hierarchy problem in physics.

We proposed also a solution of the vacuum energy density which depends of
the Tev scale and the coupling constants higher order curvature terms (in
the frame new $R^{2}$-gravity).

Throughout, units have been chosen such that $c=\hslash =1$.

Let us start with the action

\begin{equation}
S=\dint d^{4}x\sqrt{-g}[-\dfrac{\epsilon }{2}\left( v^{2}-\Phi ^{+}\Phi
\right) R+aW-\dfrac{b}{3}R^{2}-
\end{equation}

\begin{equation*}
-\frac{1}{2}\left( D_{\mu }\Phi \right) ^{+}\left( D^{\mu }\Phi \right)
-f\left( \Phi ^{+}\Phi -v^{2}\right) ^{2}]+S_{sm},
\end{equation*}

where $S_{sm}$ is a part of the action Standard Model for gauge fields and
the fermion fields.

Spin zero doublet $^{14}$ the Higgs field is $\Phi $, the Weyl term W is $%
W=R_{\mu \rho }^{2}-\frac{1}{3}R^{2}$, where $R_{\mu \rho }$\ is the Ricci
tensor and $R$\ is the scalar curvature. In the action (1) $\epsilon ,a,b$
and $f$ are dimensionless coupling constants.

\qquad The term in the action (1) $-\frac{1}{2}\epsilon v^{2}R\sqrt{-g}$\ is
the Einstein term, where we have rule $\epsilon v^{2}=M_{p}^{2}$\ and $%
M_{p}=(8\pi G)^{-\frac{1}{2}}\sim 10^{18}Gev$ is the reduced Planck mass, so
the Newton constant $G$ is not a fundamental constant. \qquad \qquad

\qquad The higher-curvature terms in the action (1) will have little effect
at low energies compared to the Einstein term. At the lowest energy, only $-%
\frac{1}{2}\epsilon v^{2}R\sqrt{-g}$ is important to the current
experimental \ \ tests of Newton's law $^{15}$ that does not contradict with
coupling constants $a\ $and $b$ have value: $a\approx b\sim 10^{60}$ . The
current experimental constraints from sub-millimeter tests to corrections of
higher-curvature terms $^{16,17}$to the Newtonian potential, give for
dimensionless constants $a$ and $b$ bounding $a$, $b<10^{62}$.

\qquad The field equations for metric $g_{\mu \sigma }$ and the Higgs field $%
\Phi $ following from the action (1) have solutions $g_{\mu \sigma
}^{(0)}=\eta _{\mu \sigma }$\ is the Minkowski metric as the metrical ground
state and nontrivial Higgs field ground state is $\left( \Phi ^{+}\Phi
\right) _{0}=v^{2}\approx (10^{3}Gev)^{2}$.

\qquad The standard way in perturbative theory one writes the metric as $%
g_{\mu \sigma }=\eta _{\mu \sigma }+\tilde{h}_{\mu \sigma }$. In the
unitarity gauge Higgs field takes the form, avoiding Goldstone bosons, we set

\begin{equation}
\ \Phi =%
\begin{pmatrix}
0 \\ 
v+\varphi%
\end{pmatrix}%
\ \ ,
\end{equation}

where the real scalar field $\varphi (x)$\ describes the excited Higgs field
connected with the Higgs particle. \ 

The part of the action (1) quadratic in the fields $\tilde{h}_{\mu \sigma }$
and $\varphi $ can be written as

\begin{equation}
\ S=\dint d^{4}x[\epsilon v\varphi R^{(1)}(\widetilde{h})+aW(\widetilde{h})-%
\dfrac{b}{3}(R^{(1)}(\widetilde{h}))^{2}-\frac{1}{2}\partial _{\mu }\varphi
\partial ^{\mu }\varphi -\dfrac{8fv^{2}}{2}\varphi ^{2}],
\end{equation}

where the Ricci tensor $R_{\mu \sigma }^{(1)}(\tilde{h})$ and the scalar
curvature $R(\tilde{h})$ can be written in a linearized form

\begin{equation}
R_{\mu \rho }^{(1)}(\tilde{h})=\frac{1}{2}(\square \tilde{h}_{\mu \rho
}-\partial _{\mu }\partial _{\sigma }\tilde{h}_{\rho }^{\sigma }-\partial
_{\rho }\partial _{\sigma }\tilde{h}_{\mu }^{\sigma }+\partial _{\mu
}\partial _{\rho }\tilde{h}),\ \ 
\end{equation}

\begin{equation}
R^{(1)}(\widetilde{h})=(\square \tilde{h}-\partial ^{\rho }\partial ^{\sigma
}\tilde{h}_{\rho \sigma })\ 
\end{equation}

and $W(\tilde{h})$ is:

\begin{equation}
W(\tilde{h})=(R_{\mu \rho }^{(1)}(\tilde{h}))^{2}-\frac{1}{3}(R^{(1)}(\tilde{%
h}))^{2}.\ \ 
\end{equation}

In the expression (3) for the fields $\tilde{h}_{\mu \rho }$ and $\varphi $
has the unwanted mixed term :

\begin{equation*}
\ \ \epsilon v\varphi R^{(1)}(\tilde{h}).
\end{equation*}%
\ \ \ \ \ \ \ \ \ \ \ \ \ \ \ \ \ \ \ \ \ \ \ \ \ \ \ \ \ \ \ \ \ \ \ \ \ \
\ \ \ \ \ \ \ \ \ \ \ \ \ \ \ \ \ \ \ \ \ \ \ \ \ \ \ \ \ \ \ \ \ \ \ \ \ \
\ \ \ \ \ \ \ \ \ \ \ \ \ \ 

We can be rid of this term making the following redefined field

\begin{equation}
\tilde{h}_{\mu \rho }=h_{\mu \rho }+\dfrac{\eta _{\mu \rho }v}{4\epsilon }%
\square ^{-1}\varphi .
\end{equation}

We find that the terms $R^{(1)}(\tilde{h})$ and $W(\tilde{h})$ take the
forms :

\begin{equation}
\ \ R^{(1)}(\tilde{h})=R^{(1)}(h)+\dfrac{3v}{4\epsilon }\varphi \ \ 
\end{equation}

and

\begin{equation}
\ W(\tilde{h})=W(h)\ 
\end{equation}

we will not keep total derivative term in eq.(9).

Putting expressions (8) and (9) in the action (3) we get the following
condition rid of the mixed term

\begin{equation}
\ \ \epsilon ^{2}=\dfrac{b}{2}\ \ 
\end{equation}

for gravitational constants $\epsilon $ and $b$. \ \ \ \ \ \ \ \ \ \ \ \ \ \
\ \ \ \ \ \ \ \ \ \ \ \ \ \ \ \ \ \ \ \ \ \ \ \ \ \ \ \ \ \ \ \ \ \ \ \ \ \
\ \ \ \ \ \ \ \ \ \ \ \ \ \ \ \ \ \ \ \ \ \ \ \ \ \ \ \ \ \ \ \ \ \ \ \ \ \
\ \ \ \ \ \ \ \ \ \ \ \ \ \ \ \ \ \ \qquad

\qquad Thus, the Planck mass $M_{p}$\ is not the fundamental scale and
depends on the coupling constant $b$\ by quadratic curvature term and the
electroweak scale $v\approx 10^{3}Gev$ which is the fundamental scale:

\begin{equation}
\ \ M_{p}=(\dfrac{b}{2})^{\frac{1}{4}}v\approx 10^{15}10^{3}Gev\sim
10^{18}Gev.\ 
\end{equation}

As a result, expression (3) has the following form

\begin{equation}
\ S=\dint d^{4}x[aW(h)-\dfrac{b}{3}(R^{(1)}(h))^{2}-\frac{1}{2}\partial
_{\mu }\varphi \partial ^{\mu }\varphi -\frac{1}{2}(8f-\frac{3}{4}%
)v^{2}\varphi ^{2}],\ \ \ 
\end{equation}

where $(8f-\frac{3}{4})v^{2\text{\ }}=m_{\varphi }^{2}$ is square mass of
the Higgs particle at $(8f-\frac{3}{4})\geqslant 0$.

\qquad Of course, the differential operator which appears in the gravity
part of action (12) is not invertible. It is necessary to add a gauge-
fixing term, in case

\begin{equation}
S_{GF}=-\dfrac{1}{2\alpha }\dint (\partial ^{\sigma }h_{\sigma \mu }\eta
^{\mu \lambda }\square \partial ^{\rho }h_{\rho \lambda })d^{4}x.\ \ \ 
\end{equation}

Going over to momentum space and using the projectors$^{1,7,18}$ for the
spin-two $P_{\mu \rho \lambda \sigma }^{(2)}$ , spin-one $P_{\mu \rho
\lambda \sigma }^{(1)}$ and the two spin-zero $P_{\mu \rho \lambda \sigma
}^{(0-s)}$\ and $P_{\mu \rho \lambda \sigma }^{(0-w)}$\ we find for actions
(12) and (13)

\begin{equation}
\ \ \widetilde{S}=S+S_{GF}=\dfrac{1}{2}\dint h^{\mu \rho }\{k^{4}[\dfrac{a}{2%
}P^{(2)}+\dfrac{1}{2\alpha }P^{(1)}-2bP^{(0-s)}+\ \ \ 
\end{equation}

\begin{equation*}
+\dfrac{1}{\alpha }P^{(0-w)}]_{\mu \rho \lambda \sigma }\}h^{\lambda \sigma
}d^{4}k.
\end{equation*}

\bigskip Then the propagator for the fields $h_{\lambda \rho \text{ }}$in
the momentum space is%
\begin{equation}
D_{\mu \rho \lambda \sigma }=\dfrac{2}{ak^{4}}P_{\mu \rho \lambda \sigma
}^{(2)}+\dfrac{2\alpha }{k^{4}}P_{\mu \rho \lambda \sigma }^{(1)}-\dfrac{1}{%
2bk^{4}}P_{\mu \rho \lambda \sigma }^{(0-s)}+\dfrac{\alpha }{k^{4}}P_{\mu
\rho \lambda \sigma }^{(0-w)}
\end{equation}

the components projectors by $P^{(1)}$ and $P^{(0-w)}$\ can be gauged away

at $\alpha \longrightarrow 0$.

Ignoring the terms proportional $\alpha $, we have for the propagator of the
momentum space is

\begin{equation*}
\ \ D_{\mu \rho \lambda \sigma }=\dfrac{2}{ak^{4}}P_{\mu \rho \lambda \sigma
}^{(2)}-\dfrac{1}{2bk^{4}}P_{\mu \rho \lambda \sigma }^{(0-s)}.\ \ \ 
\end{equation*}

In this letter we do not discuss the possible running of the coupling
constant $a,b,$and $f$\ with increasing momentum.

Let us note that redefinition (7) brings the contribution $\dfrac{v}{%
\epsilon k^{2}}$ to some vertex of the Feynman diagrams.

In this case tree-level mass of the Higgs particle is zero (which the
coupling constant $f$\ is $f=\frac{3}{32}$ ). We only have quantum
gravitational corrections in the mass of Higgs particle and the vacuum \
energy.

\qquad According to quantum corrections we have the following form of the
vacuum energy density

\begin{equation}
\rho _{vac}\simeq \dfrac{3}{4b}v^{4}<10^{-48}Gev^{4}\ \ 
\end{equation}

at $a\approx b\sim 10^{60}$ . Thus in the framework new version $R^{2}$%
-gravity with one scale, which is the electroweak scale $v\approx 10^{3}Gev$
can be also find of the solution smallness problem of the cosmological
constant.

Vacuum energy density (16) is an example of the dark energy, which counts to
about 75 per cent of the total energy density. As a result, the universe
expansion is acceleration $^{19}$.

\bigskip

\textbf{References}

1. K.S. Stelle, Phys. Rev. \textbf{D16, }953 (1977).

2. \ J. Julve, M. Tonin, Nuovo Cim. \textbf{B46,} 137 (1978).

3. \ A. Salam, J. Strathdee, Phys. Rev. \textbf{D18,} 4480 (1978).

4. \ I.G. Avramidi, A.D. Barvinsky, Phys. Lett.\textbf{\ B159,} 269 (1985).

5. \ E. Fradkin, A.A.Tseytlin, Nucl. Phys. \textbf{B201, }469 (1982);

6.\ \ A. Codello, R. Percacci, Phys. Rev. Lett. \textbf{97, }221301 (2006).

7. \ I.L. Buchbinder, S.D. Odintsov, I.L. Shapiro,

\ \ \ \ \ \ Effective action in quantum gravity,

\ \ \ \ \ \ IOP Publishing (Bristol and Philadelphia, 1992).

8. \ S. Adler, Rev. Mod. Phys. \textbf{54, }729 (1982).

9. \ A. Zee, Phys. Rev. \textbf{D23, }858 (1981).

10. A. Zee, Ann. Phys. \textbf{151, }431 (1983).

11. B.Hasslasher, F. Mottola, Phys. Lett. \textbf{95B, }237 (1980).

12. I. L. Buchbinder, S. D. Odintsov, Class. Quantum Grav. \textbf{2, }721
(1985).

13. F. David, A. Strominger, Phys. Lett. \textbf{143B,} 125 (1984).

14. S. Weinberg, Phys. Rev. Lett. \textbf{19, }1264 (1967).

15. D. J. Kapner, T. S Cook, E.G. Adelberger, J. H.Gundlalach,

\ \ \ \ \ \ B. R. Heckel, C. D. Joyle, H. E. Swanson,

\ \ \ \ \ \ Phys. Rev.Lett. \textbf{98, }021101 (2007);

16.\ \ J.F. Donoghue, Phys. Rev. \textbf{D50}, 3874 (1994);

17. \ K.S. Stelle, Gen. Rel. Grav. \textbf{9, }353 (1978).

18. \ P. van Nieuwenhuizen, Nucl. Phys. \textbf{B60, }478 (1973).

19. A. G. Riess, et al., Astrophysics J. \textbf{607, }665 (2004).

\bigskip

\end{document}